\documentclass[12pt,oneside]{article}
\usepackage{amsmath}
\input epsf.tex
\def\sect #1{\setcounter{equation}{0}}

\begin{document}
\renewcommand{\baselinestretch}{1.24} 

\title{\bf Gravitational Collapse in Constant Potential Bath}

\author{
S. Jhingan${}^{1}$\thanks{Present address: Yukawa Institute for Theoretical 
Physics, Kyoto University, 606-8502 Kyoto, Japan}~\thanks{ 
sjhingan@yukawa.kyoto-u.ac.jp}, 
N. Dadhich${}^{2}$\thanks{nkd@iucaa.ernet.in} and
P. S. Joshi${}^{3}$\thanks{psj@tifr.res.in}\\
${}^{1}$Departamento de Fisica Teorica,
Universidad del Pais Vasco,\\
Apdo. 644, E-48080 Bilbao, Spain. \\
${}^{2}$ Inter-University Centre for Astronomy \& Astrophysics,\\
Post Bag 4, Ganeshkhind, Pune-411 007, INDIA. \\
${}^{3}$Tata Institute of Fundamental Research, \\
Homi Bhabha Road, Mumbai 400005, INDIA.} 
\maketitle

\begin{abstract}

We analyse here the gravitational collapse of directed null 
radiation in a background with a constant potential such as one produced 
by a star system like galaxy in which the collapsing object is immersed. 
Both naked singularities and black holes are shown to be developing 
as the final outcome of the collapse. An interesting feature that emerges 
is that a part of the naked singularity spectrum in collapsing Vaidya 
region gets covered in the corresponding dual-Vaidya region, which 
corresponds to the Vaidya directed null radiation sitting in constant 
potential bath. The implications of such a result towards the issue of 
stability of naked singularities are discussed.

\end{abstract}


\newpage

\section{Introduction}

The spherical gravitational collapse in general relativity has been
studied extensively in recent years by several authors (see e.g.
~\cite{Magli}, and references therein). The generic picture that emerges
from these investigations is that both black holes and naked
singularities develop as a result of collapse, depending on the nature
of the initial data from which the collapse commences, for a rather
general form of matter and equation of state. Also, various regularity
conditions which are thought to be physically reasonable, are satisfied.
While the strong gravity regions forming due to collapse get necessarily
hidden behind an event horizon of gravity if a black hole forms, it is
possible for radiations and energy to escape away from such regions to a
faraway observer in the case of a naked singularity developing.

The key issue that remains open, and on which much debate is focused
currently (see e.g. ~\cite{global}), is the question of stability of such
naked singularities. The point is, even if naked singularities develop,
if they are not stable or generic in a suitable sense, they may not be
physically realizable, and a physically realistic gravitational collapse
will result in a black hole formation only. The problem however, in
making any progress towards such a conclusion, is we have no
well-defined notion of stability or genericity available presently in
gravitation theory. It is in fact known that such a formulation may
present formidable difficulties, and one may require new and
sophisticated mathematical tools towards such a purpose. Under such a
situation, a good way to make progress on this question is to subject
the currently available scenarios of collapse to physically motivated
perturbations, and then examine if the naked singularities of collapse
persist or they disappear. Actually, such studies might even help us to
generate a good definition of stability. There have been a few studies
in this direction. For example, Joshi and Krolak~\cite{nonsph} examined
quasi-spherical collapse to consider non-spherical perturbations over 
the dust collapse models. Such studies
imply that naked singularities do not necessarily go away as soon as we
depart from spherical symmetry.

Our purpose here is to investigate a somewhat different type of
perturbation over the usual scenarios considered, which is physically
motivated as we explain below. In a typical scenario for collapse a
cloud of matter distribution consisting of dust, radiation etc.
collapses onto itself under its own gravity. The exterior field to the
cloud is described by the asymptotically flat Schwarzschild solution.
The assumption of asymptotic flatness essentially means that we are
dealing with an isolated system with empty exterior. This
is of course not a realistic situation, because a collapsing star would
be a part of a star system like a galaxy. For a realistic description,
it is required to break the asymptotic flatness of the Schwarzschild
field. Since the latter is the unique solution of the Einstein vacuum
equations, this would amount to introducing some energy distribution
altering the vacuum character of the spacetime.

The question is, could this be achieved without disturbing the basic 
features of Schwarzschild field which have been observationally verified. 
The answer to this is in the affirmative. What we are seeking is the spacetime
which is for all 
observationally tested properties equivalent to the Schwarzschild field, 
yet is not asymptotically flat. The Newtonian version of the situation 
would be that the collapsing system is sitting in a cavity of constant 
potential produced by the exterior universe. In the Newtonian 
theory constant potential has no dynamics, and hence has no effect on the 
dynamics of collapse. In contrast, this is not the case for the general
theory of relativity. That is, the constant potential in the context 
of the situation under consideration is dynamically non-trivial. 
Note that the Schwarzschild solution written in terms of the potential 
has the characteristic property that it does not allow the addition of 
a constant as was allowed by the Newtonian theory. General relativity
in this sense determines the potential absolutely~\cite{nd1}. 
Without disturbing the Newtonian limit as well as the basic character 
of the field, we can add a constant to the potential which is the mark 
of the presence of exterior matter to the empty cavity embedding the 
collapsing body. It turns out that the two situations are 
electro-gravity dual of each other~\cite{nd2}. This duality is 
equivalent to the interchange of the Ricci and the Einstein tensors. 
This however does not make any difference for the Einstein 
vacuum equation but it does for the effective empty space equation for 
spherical symmetry given by $\rho = T_{ab}u^au^b = 0, \rho_n = 
T_{ab}k^ak^b = 0$ where $u_au^a = -1, k^ak_a = 0$. This admits the 
Schwarzschild solution as the general solution as the vacuum equations 
do. The dual set to these equations is $\rho_t = (T_{ab} -(1/2) Tg_{ab})u^au^b 
= 0, \rho_n = 0$, which also admits the general solution in which the 
potential attains a constant non-zero value asymptotically~\cite{nd2}. 
It is thus asymptotically non-flat but is 
rather a spacetime of constant potential. Note that, if the effective 
equation is written in terms of the Ricci components, the dual equation 
would follow from it by replacing the Ricci by the Einstein component. 
It is an interchange between the Riemann curvature and the double 
left and right dual of the Riemann. Similar to the Schwarzschild solution, 
it is possible to construct solutions dual to other solutions~\cite{nd3}, 
including the Vaidya solution.

We consider here a specific class of such models, the so called 
dual-Vaidya models, which represent the collapse of radiation shells when
the exterior is no longer vacuum Schwarzschild in the sense described
above. The paper is organised as follows. In Section 2 we discuss the
the realistic setting with the dual-Vaidya models. In Section 3 we analyze 
the causal structure of the spacetime singularity. This is followed by 
the concluding Section 4.

\section{ The realistic setting}

In the Newtonian theory the situation of spherical collapse would be 
envisioned as follows: a star or cloud of matter is collapsing in an 
empty cavity which is a part of a galaxy. Let the matter distribution 
exterior to the cavity be homogeneous, it would then produce a constant 
potential inside the cavity. The field inside the cavity would be 
described by the Laplace equation which will yield a free constant 
in its solution that could be matched with the constant potential of 
the exterior distribution. Since constant potential cannot affect the 
dynamics of collapse, it does not make any non-trivial contribution.

The situation is however different in general relativity.
Consider, for example, the Schwarzschild case. It is
asymptotically flat which means that the potential can vanish only at
infinity and nowhere else. In general relativity a  constant potential 
is thus not dynamically trivial. The Einstein vacuum equation for 
the situation under consideration ultimately reduces to the Laplace 
equation and its first integral. It is the latter that does not let 
the potential vanish anywhere else than infinity and consequently 
implying asymptotic flatness. The only
alternative to break asymptotic flatness in the most harmless way is
to relax the second equation and let the potential be given by the
general solution of the Laplace equation with two constants of
integration~\cite{nd1}. It would however amount to introduction of some
energy distribution which conforms to a string dust
distribution~\cite{let-sta} or approximates at large distance to a
global monopole~\cite{vil}. It has been shown that this modification
does not affect the Schwarzschild field appreciably, the observational
values of the physical parameters get scaled by a very small
factor~\cite{nd-kn-uy}.

It has been shown by one of us~\cite{nd2} that the
effective empty space for the situation under consideration could be
defined by the equation 
\[
\rho = T^0_0 = 0, \rho_n = T^0_0 - T^1_1 = 0.
\]
It admits the Schwarzschild (as for the vacuum equation) as the unique 
solution. Then the equation dual to it would read as ~\cite{nd2}
\[
\rho_t = T^0_0 - 1/2 T = 0, \rho_n = 0
\]
which follows from the effective equation by replacing the Einstein 
tensor by the Ricci tensor. Note that under the duality transformation, 
$\rho\leftrightarrow\rho_t, \rho_n\rightarrow\rho_n$. Here by duality we 
mean the interchange of the Ricci and the
Einstein tensor in the equation. This is termed as electro-gravity
duality for it interchanges active and passive electric parts of the
Riemann curvature~\cite{nd3}.

The remarkable thing is that the dual equation also admits the unique 
solution which is nothing but reinstating the second constant of 
integration in the solution of the Laplace equation. That is the dual 
solution to the Schwarzschild solution is not asymptotically flat and its 
potential differs from that of the Schwarzschild only by a constant. Thus 
the two solutions are absolutely equivalent at the Newtonian level. This 
is how asymptotically flatness could be avoided in the most harmless 
manner allowing us to study the collapse in the background of other 
distributions like galaxy or the rest of the Universe. This is however 
not expected to alter the basic conclusions except scaling of the crucial 
parameter which demarks black hole and naked singularity situations. In the 
same vein, the asymptotic limit of the dual Schwarzschild 
solution is a spacetime of constant potential and it is dual flat.

We take a shell of radiation collapsing onto an empty
cavity. The inner most cavity is described by the constant potential 
(dual flat) spacetime, followed by the radiation shell described by the
dual Vaidya solution and finally it is the dual Schwarzschild solution
in the exterior. Essentially, all this simply amounts to putting the
whole system into a constant potential background. Since this is
non-trivial in GR, it would be worthwhile to study its effect on the
process of gravitational collapse.

The spacetime under consideration can be divided into three regions I, II
and III, corresponding to the dual flat, the dual Vaidya
and the dual Schwarzschild spacetime, respectively. 

The dual vacuum spacetime (region I) can be conveniently written in the 
$(V,R,\theta,\phi)$ coordinates as
\begin{equation}
ds_I^2 = -(1+k)dV^2 + 2dVdR + R^2 (d\theta^2 + sin^2\theta d\phi^2)
\label{dvacuum}
\end{equation}
here null coordinate $V$ denotes the advanced time and constant negative
$k$ characterizes the strength of the constant field 
resulting from the dual vacuum in the limit of vanishing mass. 
The range of $k$ is restricted to $-1<k\leq0$, in order to preserve the 
signature of the metric, as otherwise the metric becomes degenerate. 
The only nonvanishing component of the stress-energy tensor is 
$$
T^{v}_{v} = -\frac{k}{R^2} .
$$
The weak as well as strong energy conditions are satisfied in the 
spacetime.

We note that there is a mild singularity at the center $r=0$ in the dual 
Minkowski spacetime above. However, as will be clear, it is not a genuine 
physical feature in any sense because $\sigma^{2}R_{ij}K^iK^j$ goes to zero 
in the limit, $\sigma \rightarrow 0$, of approach to the singularity 
(here $K^i$ is the tangent vector to nonspacelike geodesics, and 
$\sigma$ is the affine parameter along the same). Hence no 
radial test particles (ingoing or outgoing) feel any gravitational tidal
forces. The initial data is regular because there are no trapped surfaces
on any spatial slice below the genuine curvature singularity at $v=0$, 
$r=0$, the first point of the dual Vaidya Spacetime. In the Newtonian
limit the spacetime corresponds to a constant potential which 
understandably does not give rise to any physical pathology. 

Region II consists of a dual Vaidya spacetime, by which we mean an
imploding radiation field over a background constant potential field
generated by $k$. It can be given in general by a metric of the form
\begin{equation}
ds_{II}^2 = -(1-\frac{2g}{r})dv^2 + 2dvdr + r^2 (d\theta^2 + 
sin^2\theta d\phi^2) .
\label{dvaidya}
\end{equation} 
Here $g(v,r)$ is interpreted as the  mass function, which
can be defined in general in spherically symmetric spacetimes, as
\begin{equation}
g(v,r) \equiv \frac{r}{2}{R}^3_{232}=\frac{r}{2}
(1-g^{\mu \nu}\partial_{\mu}r\partial_{\nu}r),
\label{mass}
\end{equation}
where $r$ is the areal radius and ${R}^3_{232}$ is 
the component of the Riemann-Chritoffel tensor. 
It is a positive definite quantity due to the positivity of matter 
density.

As $r$ is a coordinate here we have
\[
g(v,r) = \frac{r}{2} (1-g^{rr}).
\]
The null coordinate $v$ is defined for range $v\geq 0$. Regions $I$ and
$II$ are matched across $v=0$ null hypersurface where radiation starts
to collapse. For a realistic evolution it is necessary that there is no
flux of radiation across the $v=0$ null line, and that the mass function
attains the same value at it when approached from either side.
From equation~(\ref{mass}) the above stated matching
implies following condition on the mass function,
\begin{equation}
g(v,r) |_{v=0} =-\frac{kr}{2} .
\label{match}
\end{equation}
For simplicity we work throughout with a source of radiation given by a
linear function such that
\begin{equation}
2g(v,r) = \lambda v - kr .
\label{eff-mass}
\end{equation}
(Note that the choice above ensures matching of mass function along $v=0$
null hypersurface).
The only nonvanishing components of the stress-energy tensor are 
$$
T^{v}_{v} = -\frac{k}{r^2} , T^{r}_{v} = \frac{\lambda}{r^2} .
$$
The weak as well as strong energy conditions are satisfied in the 
spacetime because we have $\lambda >0$ and $k<0$.

The source of radiation is switched off at $v=T$ and the spacetime (Region
II) is here joined with a dual Schwarzschild spacetime (Region III) 
\begin{equation}
ds_{III}^2 = -(1-\frac{\lambda T}{r}+k)dv^2 + 2dvdr + r^2 (d\theta^2 + 
sin^2\theta d\phi^2) 
\label{dsch}
\end{equation} 
with mass $M = (\lambda T - kr)/2$ (see Fig. 1).

\vglue0.5cm

\hglue 5 cm {\epsfysize=8cm\epsffile{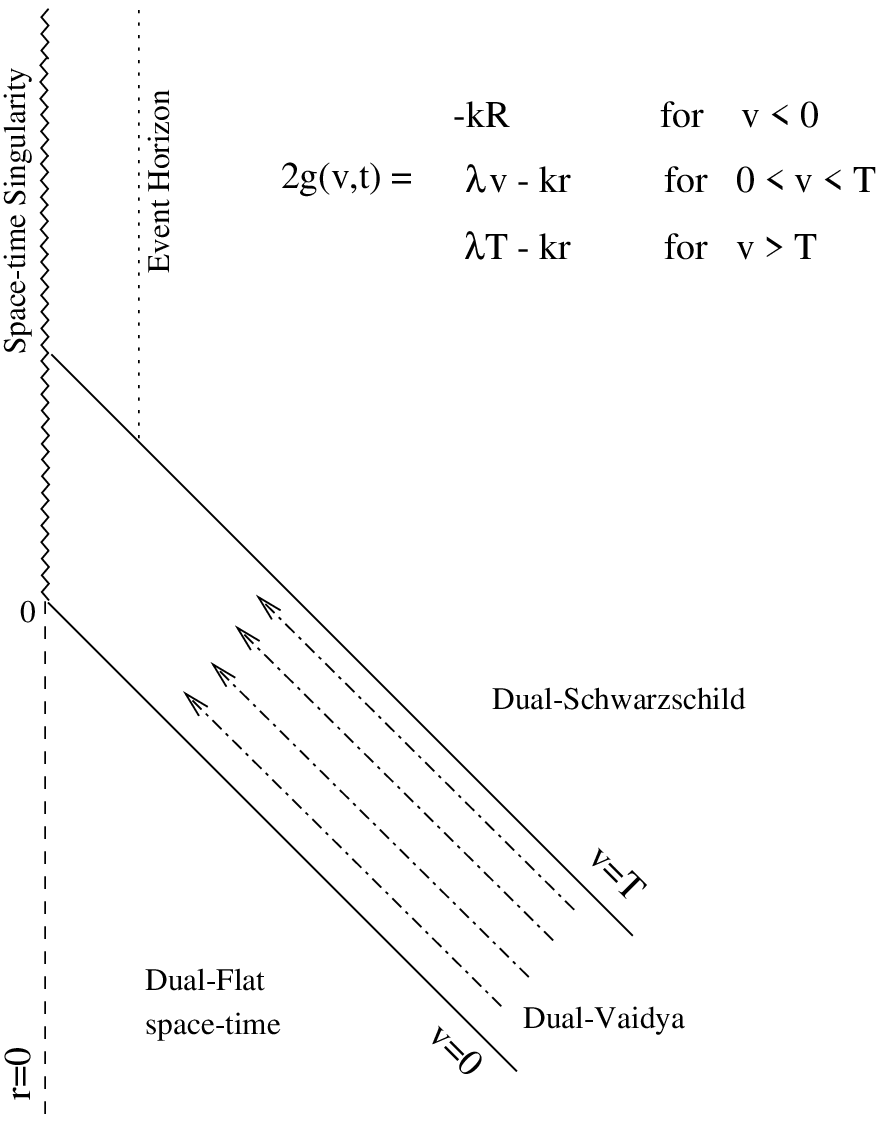}}

\vglue 0.5cm

{\centerline{{\bf Figure 1} Gravitational collapse of radiation shells}}

\vglue 0.5 cm

\section{Nature of the final singularity}

Earlier investigations of collapse models with a linear mass function were
done by Kuroda~\cite{Kuroda} and Papapetrou~\cite{Papapetrou} where they
showed the existence of naked singularity for weak enough radiation
collapsing along $v=$ constant lines. Radial null geodesics were analysed 
by Papapetrou clarifying the null structure of spacetime. The cases with
more general mass function were analysed by Rajagopal
and Lake~\cite{Rajagopal} and others~(\cite{global}). Also it is well-
known that singularities arising in these spacetimes are strong
curvature singularities.

For our purpose here we have restricted the discussion to the choice of
a linear mass function (\ref{eff-mass}). This has the advantage 
of a clearer physical 
interpretation in that it relates the outcome of collapse as black
hole or a naked singularity in terms of rate of collapse or mass implosion 
at the center. The key idea of the analysis
here is that the nature of monopole field maintains the self-similar
nature of the spacetime metric and we could use the analysis developed
earlier (see, e.g. chapter 6~\cite{global}). The
dual-Vaidya region admits a homothetic killing vector
\[
\xi = v\left(\frac{\partial}{\partial v}\right) + 
r\left(\frac{\partial}{\partial r}\right)
\]
which is given by the Lie derivative
\begin{equation}
{\cal L}_{\xi}g_{\mu\nu}= 2g_{\mu\nu}
\label{geo0}
\end{equation}
Define now $K^{\mu}=dx^{\mu}/ds$ as tangent to non-spacelike geodesics,
where $s$ is an affine parameter. Then,
\begin{equation}
g_{\mu\nu}K^{\mu}K^{\nu} = B
\label{geo1}
\end{equation}
where $B=0$ for null and $+1 (-1)$ for spacelike (timelike) vectors.
Along non-spacelike geodesics we have
\begin{equation}
\frac{d}{ds}(\xi^{\mu}K_{\mu}) = \xi_{\mu;\nu}K^{\mu}K^{\nu} 
\label{geo2}
\end{equation}
which can be integrated, after using eqns~(\ref{geo0}) 
and~(\ref{geo1}) to reduce the expression on the right hand 
side of equation above to $B$, as
\begin{equation}
\xi^{\mu}K_{\mu}=Bs+C
\label{geo4}
\end{equation}
where we have used $K^{\mu}_{;\nu}K^{\nu}=0$ and $C$ is the integration
constant. From the expression for Killing vector we get from
~(\ref{geo4})
\begin{equation}
rK_{r}+vK_{v}=Bs+C .
\label{geob}
\end{equation}
 
Now defining $P(v,r)$ as
\begin{equation}
K^{v}=\frac{P(v,r)}{r}=\frac{dv}{ds}
\label{geo5}
\end{equation}
and using eqn.~(\ref{geo1}) we get
\begin{equation}
K^r = \left(1-\frac{2g}{r}\right)\frac{P}{2r} + 
\frac{r}{2P}\left(B-\frac{l^2}{r^2}\right) .
\label{geo6}
\end{equation}

Writing~(\ref{geo1}) explicitly and using eqn~(\ref{geob}) for $K_v$
we have
\begin{equation}
r^2 [X+kX-\lambda X^2 -2](K_{r})^2 +2r(Bs+C)K_r+X(l^2-Br^2) = 0,
\label{geo7}
\end{equation}
where we have defined the similarity variable $X=v/r$. 
Eqn~(\ref{geo7}) can be solved for $K_r$ as
\begin{equation}
rK_r = \frac{(Bs+C)\mp\sqrt{(Bs+C)^2+
X(l^2-Br^2)(2+\lambda X^2 - X - kX)}}{2+\lambda X^2 - X - kX}
\label{geo8}
\end{equation}
and $K_v$ can be obtained using~(\ref{geob}). Since $K^v=g^{vr}K_r$ we
get the expression for $P$, defined in~(\ref{geo5}), 
using~(\ref{geo8}) as
\begin{equation}
P(v,r) = \frac{C(1+As)\mp C \sqrt{(1+As)^2+
X[L^2-(Ar^2/C)](2+\lambda X^2 - X - kX)}}{2+\lambda X^2 - X - kX}
\label{geo9}
\end{equation}
where $A=B/C$ and $L=l/C$.

The point $v=0, r=0$ is the first point on the singularity curve. To
analyse nature of this point one can try to analyse the outgoing
singular geodesics terminating at the singularity in the past. We
consider a parameterization ($s$) of these outgoing singular geodesics
in such a way that in the limit $s\rightarrow0$ we approach singular
point $(0,0)$. Let
\begin{equation}
\lim_{s\rightarrow0} X = \lim_{s\rightarrow0}
\frac{v}{r} = X_0  ,
\label{root1}
\end{equation}
along a singular geodesic.
Using eqns.~(\ref{geo5}),~(\ref{geo6}) and~(\ref{geo9}) in
eqn.~(\ref{root1}) above we get
\begin{equation}
X_0=\lim_{s\rightarrow0}
\frac{dv/ds}{dr/ds}= \frac{2[1+Q(X_0)]^2}{(1-\lambda X_0+k)[1+Q(X_0)]^2 -
L^2(2+\lambda X_0 -kX_0-X_0)^2}
\label{root2}
\end{equation}
where $Q(X)= [(1+As)^2+X[L^2-(Ar^2/C)](2+\lambda X^2 - X - kX)]^{1/2}$
and $r=r(X)$.

Since we want to analyze the effect of constant potential term $k$ on
the final state of collapse, it is sufficient for our purpose to analyse
the simple case of radial null geodesics, i.e., $L=0$ and $B=0$.
Eqn.~(\ref{root2}) simplifies to
\begin{equation}
X_0= \frac{2[1+Q(X_0)]^2}{(1-\lambda X_0+k)[1+Q(X_0)]^2}
\label{root3}
\end{equation}
and hence, on further simplification,
\begin{equation}
X_0=a_{\pm}=\frac{(1+k)\pm \sqrt{(1+k)^2-8\lambda}}{2\lambda} .
\label{finalroot}
\end{equation}
Therefore, real values of $a_{\pm}$ are possible only if
$\lambda \leq (1+k)^2/8$ (See Fig. 2). 
 
\vglue0.5cm

\hglue 2 cm {\epsfysize=8cm\epsffile{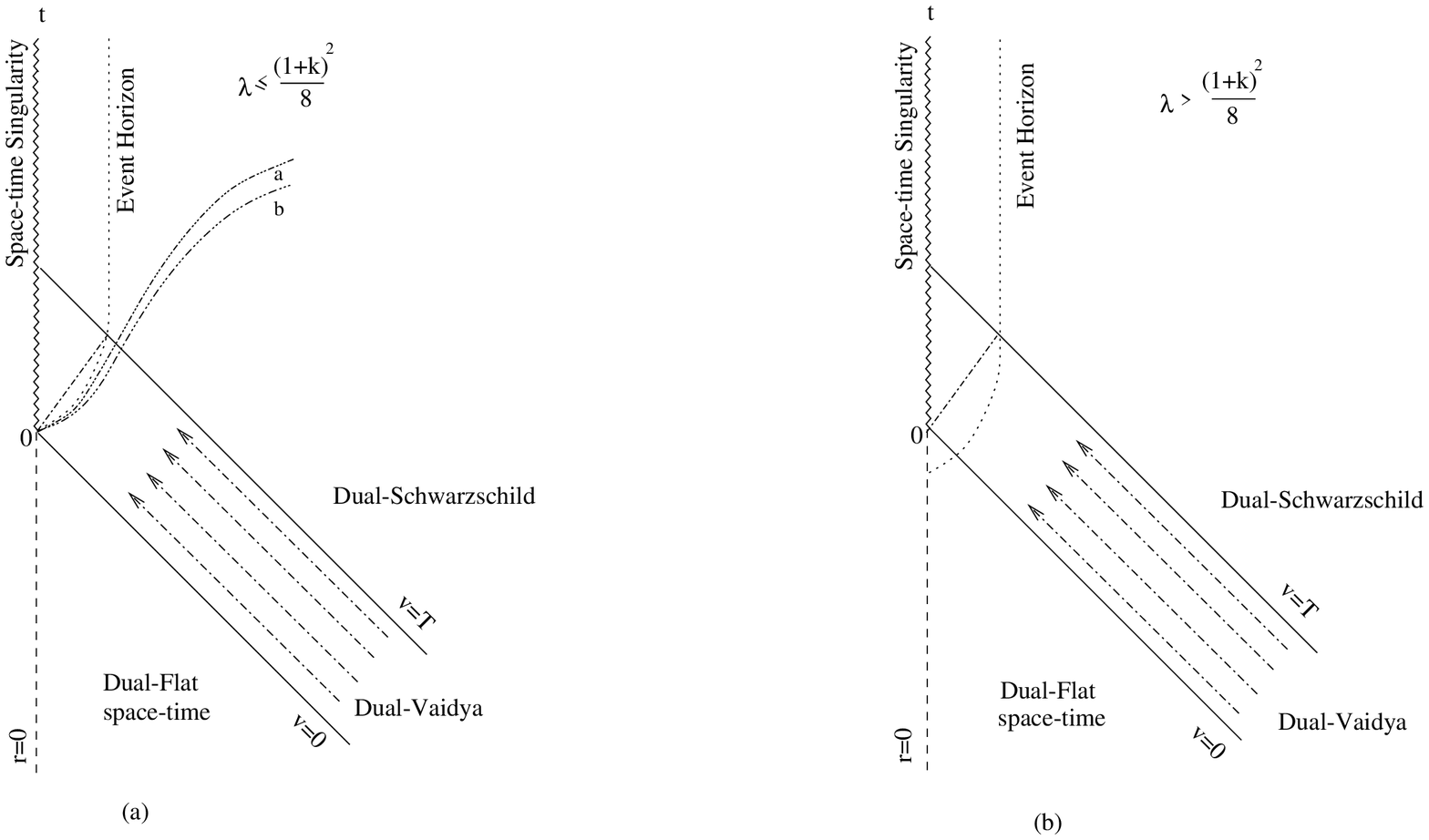}}

\vglue 0.5cm

{\centerline{{\bf Figure 2} Final singularity in gravitational collapse}}

\vglue 0.5 cm

The singularity arising in the Vaidya-Papapetrou model was shown 
to be a strong curvature singularity, and also a scalar polynomial
singularity~\cite{global}. In fact, all known cases of strong 
curvature singularities in Einstein equations are also scalar
curvature singularities. In the model here also the singularity 
continues to be a strong curvature singularity, as expected, with
the Kretschmann scalar 
\[
{\cal K} = R^{ijkl}R_{ijkl} = 4\frac{k^2}{r^4} - 8\frac{k\lambda X}{r^4} 
+ 12\frac{{\lambda}^2 X^2}{r^4}
\]
which always diverges at the singularity. It is interesting to note 
that the curvature scalar ${\cal R}\equiv g^{ij}R_{ij}$ and the Ricci 
scalar $R^{ij}R_{ij}$ also diverge in the limit of approach to
the singularity,
\[
{\cal R} = -2\frac{k}{r^2},  R^{ij}R_{ij} = 2\frac{k^2}{r^4}
\]
which were vanishing in the Vaidya-Papapetrou case.

One can further measure the rate of curvature growth along the 
singular geodesics which come out, in the limit of approach to the 
singularity. The result is,
\[
\lim_{s \rightarrow 0} s^2 R_{ij} V^i V^j = \frac{4 \lambda}{(1+k)}
\]
where $s$ is the affine parameter, and we restrict to outgoing 
radial $(l=0)$ null $(B=0)$ geodesics.
We work in the parameterization $s=0$ corresponding to the first point on 
the singularity curve. It is thus seen that the singularity continues 
to be a strong curvature singularity. Again, setting $k=0$ gives us the 
earlier results in the Vaidya models.

\section{Concluding remarks}

We studied here the Vaidya collapse with a monopole field and
examined the formation of black holes and naked singularities.
Our purpose was to examine how the perturbation induced by the external
matter fields affect the formation or otherwise of the naked
singularity. The relevant question is whether the effect of such
external fields could remove the occurrence of the same.
We find that a part of the range of the parameter space gets
covered once the external potential is taken into account.

This scenario is motivated by the fact that in the real Universe 
a collapsing object is always sitting in a cavity which is enclosed by the 
other matter distribution like a galaxy, cluster of galaxies, rather the 
rest of the Universe. Strictly speaking, asymptotic flatness character of 
the Schwarzschild solution does not permit the existence of any other 
matter in the Universe. To make the setting accord with the non-empty 
rest of the Universe and Mach's principle, we need to break the 
asymptotic flatness of the Schwarzschild solution ~\cite{nd5}. The 
constant $k$ is the measure of the uniform potential produced by the rest 
of the stars in the galaxy in the empty cavity enclosing the collapsing 
system. That is why it would be negative and its measure would be of the 
order of {\it O}($10^{-6}$) for the star velocity of 200 km/sec. It 
would therefore not disturb any observations materially.

It is worth noting however, that since the condition $|k|<1$ is always 
respected, there would {\it always} (even for large $k$) exist a 
nonzero measure range of parameter space available for which the 
naked singularity definitely develops. In that sense, the potential 
generated by the exterior universe 
does not remove the naked singularity, which thus displays
stability with respect to this particular mode of perturbation. 
Since there is no general definition available to check the
stability of naked singularities as indicated earlier, the only
way available is to check the stability with respect to classes of
perturbations which are essentially physically motivated
(see e.g. ~\cite{reza}). It would be reasonable to subject various 
classes of collapse models resulting into naked singularities 
to such physical perturbations to check for the stability 
(see e.g. ~\cite{jhingan97} for a detailed discussion on dust models 
and how initial data is related to the formation of naked singularities 
and black holes). This may be perhaps the only possible way available 
towards a suitable formulation of the cosmic censorship hypothesis.

\section*{Acknowledgements}
SJ would like to thank the support from the Basque Government 
Fellowship. This work was supported partially by the University of Basque 
Country grant UPV172.310-G02/99.  ND would also like to thank Alberto 
Chamorro and Jose Senovilla for their warm hospitality.  

\section*{Appendix}

The geodesic equations, for the spacetime metric~(\ref{dvaidya})
in region II, are written as
\begin{equation}
\frac{d}{ds}[r^2sin^2{\theta}K^{\phi}]=0
\label{phi}\tag{A1}
\end{equation}
\begin{equation}
\frac{dK^{\theta}}{ds}+\frac{2}{r}K^r K^{\theta} - sin\theta cos\theta
(K^{\phi})^2 = 0
\label{theta}\tag{A2}
\end{equation}

\begin{align}
\frac{dK^r}{ds} + 2\left(\frac{g}{r^2}-\frac{g_r}{r}\right)K^rK^v 
+\left[\frac{g_v}{r} -\left(\frac{g}{r^2}-\frac{g_r}{r}\right)\left(1-
\frac{2g}{r}\right)\right](K^v)^2  \notag \\
- \left(1-\frac{2g}{r}\right)
r[(K^{\theta})^2+ sin^2{\theta}(K^{\phi})^2] = 0 
\label{thatsit}\tag{A3}
\end{align}

\begin{equation}
\frac{dK^v}{ds} + \left(\frac{g}{r^2}-\frac{g_r}{r}\right)(K^v)^2 -
r[(K^{\theta})^2+ sin^2{\theta}(K^{\phi})^2] = 0 .
\label{halfnull}\tag{A4}
\end{equation}
Here superscripts denote corresponding partial derivatives.
Integrating eqns. (\ref{phi}) and (\ref{theta}) gives solution of the form
\begin{equation}
K^{\theta}=\frac{lsin{\beta} cos{\phi}}{r^2}, \ \ \
K^{\phi}=\frac{lcos{\beta}}{r^2sin^2{\theta}}
\label{solution}\tag{A5}
\end{equation}
where $l$ is impact parameter and $\beta$ is an isotropy parameter
satisfying $sin{\phi}tan{\beta}=cot{\theta}$.

\end{document}